\documentclass[sigconf,anonymous=false, nonacm]{acmart}
\pdfoutput=1
\newcommand{\mohak}[1]{{\color{red} [MG: {#1}]}}
\usepackage{enumitem}
 \usepackage{balance}
\newtheorem{theorem}{Theorem}[section]
\newtheorem{corollary}[theorem]{Corollary}

\newtheorem*{lemma*}{Lemma}
\newtheorem{example}[theorem]{Example}
\newtheorem*{example*}{Example}
\newtheorem{definition}[theorem]{Definition}
\newtheorem*{theorem*}{Theorem}
\newtheorem{observation}{Observation}
\newtheorem{assumption}{Assumption}
\newtheorem*{proposition*}{Proposition}

\begin{document}

\title{Pricing Personalized Preferences for Privacy Protection in Constant Function Market Makers}
\author{Mohak Goyal}
\affiliation{%
  \institution{Stanford University}
  \city{Stanford}
  \state{CA}
  \country{USA}
  \postcode{94305}
}
\email{mohakg@stanford.edu}

\author{Geoffrey Ramseyer}
\affiliation{%
  \institution{Stanford University}
  \city{Stanford}
  \state{CA}
  \country{USA}
  \postcode{94305}
}
\email{geoff.ramseyer@cs.stanford.edu}

\thanks{The authors thank 
%\geoff{geoff: adding myself here as thanking them} 
Ashish Goel, Sahasrajit Sarmasarkar, and Karan Chadha for helpful discussions. This work was funded by the Future of Digital Currency Initiative (FDCI), Stanford University, and
the Stanford IOG Research Hub.}
\newcommand{\geoff}[1]{\begingroup \color{red} #1 \endgroup}

\begin{abstract}
 Constant function market makers (CFMMs) are a popular decentralized exchange
 mechanism and have recently been the subject of much research, but major  CFMMs give traders no privacy. Prior work 
 proposes randomly splitting and shuffling trades to give some privacy to \textit{all} users \cite{chitra2022differential}, or
 adding noise to the market state after each trade and charging a \textit{fixed} `privacy fee' to all traders \cite{frongillo2018bounded}. In contrast, we propose a noisy CFMM mechanism where users specify personal privacy requirements and pay personalized fees. 
 We show that the noise added for privacy protection creates \textit{additional arbitrage} opportunities. 
  We call a mechanism \textit{priceable} if there exists a privacy fee
  that always matches the additional arbitrage loss in expectation. We show that a mechanism is priceable if and only if the noise added
  is zero-mean in the asset amount.
  We also show that priceability and setting the right fee are necessary for a mechanism to be  \textit{truthful}, and that this fee is
  inversely proportional to the CFMM's liquidity. 
\end{abstract}


\maketitle

\section{Introduction}

Market makers are market participants who facilitate the quick exchange of assets by offering to both buy and sell an asset at any given time. 
In doing so, they take the risk of holding an asset whose price may change, and in exchange, they charge a fee on each transaction or make a `bid-ask spread'. 
While market making has been a long-studied topic in finance
\cite{amihud1980dealership,kyle1985continuous,glosten1985bid}, 
a simple market-making strategy called Constant Function Market Makers (CFMMs) \cite{angeris2020improved,goyal2022finding,mohan2022automated} has  risen to prominence in decentralized finance (DeFi), facilitating billions of USD in daily trade volume. 
CFMMs are equivalent to the cost-function-based market makers (as illustrated in, e.g., \cite{frongillo2023axiomatic}) of
prediction markets \cite{hanson2007logarithmic,chen2012utility}.

A CFMM maintains non-negative reserves $(x,y)$ of two assets $X$ and $Y$ ( henceforth called its ``state''), provided by a \emph{liquidity provider (LP)}. A CFMM trades according to an eponymous \emph{trading function} $f(x,y)$ of its state. It accepts a trade $(\Delta x, \Delta y)$ from reserves $(x,y)$ to $(x + \Delta x, y - \Delta y)$ if and only if $f(x + \Delta x, y - \Delta y) = f(x, y)$.  Assumption~\ref{ass:cfmm} gives technical conditions on $f$.

%The loss of the LP relative to the counterfactual strategy of ``buy-and-hold'' is called the ``divergence loss'' \cite{xu2023sok}. Part of the divergence loss is \textit{inventory costs} and can be hedged via options \cite{chitra2022note}. The unavoidable part, which is due to \textit{arbitrage}, is called the ``loss-vs-rebalancing'' or LVR \cite{milionis2022automated}. 

For simplicity, we refer to $X$ as a volatile asset and $Y$ as a stable numeraire, and $p$ as the price of $X$ given in $Y$. The price of an infinitesimal trade on the CFMM is called its ``spot price'' (Definition~\ref{defn:spot}).

One drawback of CFMMs is that they do not preserve traders' privacy. We refer to a trader's information as the amount of $X$ they buy or sell. A CFMM publicly quotes prices as a function of trade size. As a result,
observing these prices before and after a trade reveals the trade amount precisely due to the curvature of the CFMM trading function~\cite{angeris2021note,chitra2022differential}. This implies that hiding CFMM state via cryptographic 
tools 
%encryption, such as Zero-Knowledge Proofs (ZKPs), 
cannot provide privacy on individual trades \cite{whi20}.
\citet{angeris2021note} formalize this idea and show that the black-box use of trusted hardware or cryptographic primitives with CFMMs is unlikely to be fruitful for privacy protection. We adopt their model of the adversary and its attack capabilities (\S~\ref{sec:attack}).

\citet{chitra2022differential} argue that differential privacy (DP) is a natural notion for studying CFMM privacy. DP is a framework to collect users' data, protecting individual privacy while approximating aggregate statistics.  In a CFMM, this idea translates to aggregating individual trades with privacy protection while approximating prices. 
%The domain of CFMMs is forgiving in that 
If the differentially private CFMM mechanism's price deviates significantly from the `true' price, arbitrageurs are incentivized to correct the CFMM's price and realign it with the true price.

\subsection{Our Contributions}
We adopt the \textit{personalized local differential privacy} (PLDP) framework (Definition~\ref{def:pldp}) of \citet{chen2016private} for CFMMs. A user $i$ with a trade of selling $\Delta_i$ units of X ($\Delta_i < 0$ implies buying X) specifies their desired privacy level $\varepsilon_i \in \mathbb{R}_{\geq 0}$ (lower $\varepsilon_i$ implies better privacy) and a masking interval $\tau_i = [l_i,u_i]$ to mask their trade within, such that $l_i, u_i \in \mathbb{R}$ and $\Delta_i \in \tau_i$.  %The precise definition is in \S\ref{sec:prelim} (Definition~\ref{def:pldp}). 
%This is a generalization of the local differential privacy (LDP) framework (Definition~\ref{def:ldp}) \cite{warner1965randomized,kasiviswanathan2011can} where each user has the same $\tau$ and $\varepsilon$.
 While we adopt the privacy notion of \citet{chen2016private}, our motivation is different -- they use the model for spatial data collection and study the error of aggregate statistics. In contrast, we use it to mask trades in CFMMs and study the \textit{economic implications}. %Same as most LDP-preserving methods, 
 We study mechanisms that add noise to the market state with each user's data separately.

Privacy is not free. In the context of data collection, it causes error in the aggregate statistics, and in the context of CFMMs, it creates \textit{additional} arbitrage opportunities.%\footnote{We do not aim to tackle the loss due to arbitrage faced by traditional CFMMs. There are papers on designing trading fee mechanisms for compensating the LPs for the arbitrage loss. See, for example, \citet{milionis2022automated,fritsch2021note,evans2021optimal}.}
 % This paper finds the \textit{privacy fee} amount that adequately compensates the LP for the \textit{additional} arbitrage created by noise-adding mechanisms for privacy protection on CFMMs. We do not investigate the best way to integrate these two fees, but charging the sum of the two is a natural choice, which is what we adopt here.}.
Consider a simple example. 
At a point in time, a CFMM's spot price is $p$. The price on an external market (with in infinite liquidity, say, an off-chain exchange) is $\hat p > p$. Then an arbitrageur can buy some $X$ from the CFMM and sell it to the external market, raising the CFMM's spot price to $\hat p$ under the `optimal' arbitrage \cite{angeris2020improved}.
If the CFMM uses a noise-adding privacy protection mechanism and has non-zero curvature, then its spot price is distorted from $\hat p$ for the privacy of the trade. Then, the arbitrager has an \textit{additional} arbitrage opportunity to profit off the CFMM by moving the spot price again to $\hat p$. 
To be sustainable, the CFMM must charge an additional fee for privacy preservation to account for this additional arbitrage. However, finding this fee amount without knowing $\hat p$ is not trivial. %We refer to the property of the noise-adding mechanisms where it is possible to calculate the fee amount that exactly pays for the additional arbitrage loss, without knowing $\hat p$, as its \textit{priceability} (Definition~\ref{def:priceability}). 

We call a private CFMM mechanism \textit{priceable} (Definition~\ref{def:priceability}) if there exists a fee,
independent of the true price,
that pays for the additional arbitrage.
%When there exists a finite fee amount that exactly pays for the additional arbitrage without knowing the true price $\hat p$, we say that the private CFMM mechanism is \textit{priceable} (Definition~\ref{def:priceability}).
We show that a noisy CFMM mechanism is priceable if and only if the noise is zero-mean in the asset amount. The proof is technical, but intuitively, a zero-mean noise does not create additional liquidity in expectation. Pricing the additional arbitrage at the post-trade spot price ensures that an arbitrageur does not gain in expectation, regardless of the true price.
% This does not need the knowledge of the true price. %Zero-mean noise is used widely in DP since it leads to an unbiased mean estimate. In our setup, it has another benefit of facilitating priceable mechanisms for privacy in CFMMs.

We also define \textit{truthfulness} of a noise-adding CFMM mechanism, such that a trader without innate privacy requirements has the incentive to leave the CFMM spot price at the external market price $\hat p$ in one trade (Definition~\ref{def:truthful}). We show that zero-mean noise distribution is necessary for truthfulness. With the right privacy fee, a zero-mean noise distribution leads to a truthful mechanism. This ensures that the noisy CFMM's spot price will closely follow the price on an external exchange as long as arbitragers are present.

For priceable mechanisms, we formulate the privacy fee as a function of the CFMM state, the trading function, and the trader's privacy requirements. We show that a more liquid (Definition~\ref{def:liquidity}) CFMM requires a smaller privacy fee. In fact, the privacy fee of a trade is inversely proportional to the liquidity (Proposition~\ref{thm:liq-fee}). This aligns with the fee-liquidity relationship of \citet{frongillo2018bounded} in an information aggregation context.% with a constant fee and the same privacy guarantees for all users.%; they designed a \textit{bounded-loss} market for information aggregation.

%We then do numerical simulations to study the price impact of privacy protection on the market and find that for reasonable trade patterns, this impact is low. We also study the volatility of CFMM spot prices and find that it is not much different from that of a non-private CFMM.
\subsection{Related Work}

\subsubsection{Prediction Markets} ~

Closest to our work is that of \citet{frongillo2018bounded}, which gives a bounded-loss differentially-private cost-function-based prediction market via a similar mechanism---adding noise to the market after each trade. Our work differs in the following ways:
\begin{itemize} [leftmargin = 10pt]
\item They mandate the same privacy specification $(\tau,\varepsilon)$ and privacy fee for all traders regardless of trade size. Further, they restrict the maximum size of a trade.
We give a framework for specifying personalized privacy requirements in $\varepsilon_i$ and $\tau_i$ for trader $i$, and find a personalized privacy fee, which decreases with $\varepsilon_i$ and increases with widening the masking interval. Thus, we do not restrict the trade size, except when it causes the CFMM to run out of an asset.

%    \item Their premise has no ``external market price'' and the price resulting from the trades on their market must reflect the information aggregated from the trades. In this setup, it is important to limit the amount of noise the mechanism adds to be able to get a more `precise' estimate of the aggregate information. On the other hand, we assume the existence of an external market price and that arbitragers can re-align the CFMM's spot price to it.% if it deviates significantly from the external market price. 

    \item They use the continual observation technique of \citet{dwork2010differential} and \citet{chan2011private} wherein old noise drawings are re-used cleverly. The benefit of this approach is that the total noise over $T$ trades is limited to $O(\log T)$ (informally, many noise terms cancel each other out). However, the drawback is that the noise added after each trade is also $\Omega(\log T)$. Such a mechanism is not directly useable for CFMMs in DeFi since (1) the total number of trades, $T,$ is often not fixed, (2) storing $O(T)$ noise drawings in a hidden manner for future re-use may not be possible, (3) the privacy fee which accounts for the additional arbitrage would be $\Omega(\log T)$ larger since it depends on the magnitude of noise added after individual trades.

    We do not aim to reduce the total noise variance since, per our assumptions, arbitragers keep the CFMM price close to accurate. In exchange, we use smaller noise variance on individual trades to facilitate a smaller personalized privacy fee.
    
   % \item Their fee mechanism ensures that the loss of the market maker is bounded in the worst case, whereas our fee mechanism pays for the privacy costs \textit{exactly}.

\end{itemize}

Earlier, \citet{cummings2016possibilities} showed that without a fee, noise-adding bounded-loss cost-function-based market makers must have a `fast'-growing $\varepsilon$ as more trades are made by the market (i.e., quickly diminishing privacy guarantees).

\subsubsection{Decentralized Finance}~

\citet{chitra2022differential} give an algorithm for splitting and shuffling trades in a block to ensure privacy. They study the worst-case price discrepancy between their algorithm and a non-private CFMM, and its trade-off with DP guarantees. Our paper differs in the following:
\begin{itemize}  [leftmargin = 10pt]
\item They consider blockchains with ``consensus rules for executing trades in a particular order.'' In comparison, our mechanism does not require this capability. Most mainstream blockchains do not have this capability. We, however, require the blockchain to have verifiable randomness, the same as their mechanism.
\item Their mechanism provides the same privacy guarantees to all traders, whereas we provide a personalized mechanism.
\item Their shuffling mechanism requires many trades in a block to be able to \textit{hide} trades, and their privacy guarantees depend on the number of trades in a block. While this is often true, having a mechanism invariant to the number of trades in a block, such as ours, is useful for some contexts.
\end{itemize}
%\cite{angeris2021note} argued that cryptographic encryption will be insufficient to preserve privacy in CFMMs and gave a model of an adversary -- we use their model in this paper,

Outside of CFMMs, \citet{zswap} use batch auctions and homomorphic encryption to enable private swaps in DeFi. 
%\geoff{They use homomorphic encryption, not ZK, if I recall correctly.Their thing is to batch trades together and then only reveal the aggregate (and IIRC, they leave differential privacy of trades as an open problem in their whitepaper).  This framing sets it up as too similar to this work, I think.}
\citet{davidow2023privacy} designed a verifiable LDP system for payments. See \citet{dai2021flexible,zhu2022data,zhou2022sok} for more perspectives on privacy in DeFi.

\subsubsection{Differential Privacy}~

We adopt the personalized local differential privacy (PLDP) framework of
\citet{chen2016private}. The original framework was motivated by a model of users sharing spatial data and requiring the ability to mask their trade in a specified region. For example, user $A$ may not mind disclosing that they are in New York City but do not want to share the location within New York City. Another user, $B$, may be willing to disclose that they are in Ghana, but not any more. This framework is better suited for masking trade sizes than simple LDP. The masking interval desired by a trader buying $1$ unit $X$ would generally be different from that desired by a trader selling $100$ $X$.

See \citet{dwork2010differential} for an overview of DP. See \citet{yang2020local} for a survey of the applications of LDP, which include Google Chrome Browser \cite{erlingsson2014rappor} and Apple OS~\cite{macos}.

\section{Preliminaries}
%We here discuss the basic concepts of CFMMs and DP.
\subsection{Constant Function Market Makers (CFMMs)}
A CFMM is an automated market-maker 
parameterized by a trading function $f$.  We focus the scenario where a CFMM trades between a volatile asset $X$ and a stable numeraire asset $Y$, which is the context in which CFMMs are most often employed.
A liquidity provider endows the CFMM with nonnegative 
\textit{reserves} $(x,y)\in \mathbb{R}^2_+$ of each asset.
%Generally, a CFMM can trade in any number of assets. However, most commonly deployed CFMMs trade in only two assets, which is the case we study in this paper.  
%At any time, a liquidity provider holds 
% non-negative amounts $(x,y) \in \mathbb{R}^2_+ $ of assets $X$ and $Y$ in the CFMM's \textit{reserves}.
A CFMM with reserves $x,y$ and trading function $f$ accepts any trade $(\Delta_x, \Delta_y)$ that results in reserves $x + \Delta_x, y - \Delta_y$ 
if and only if $f(x+\Delta_x,y -\Delta _y) = f(x,y)$. 
%If the trade is accepted, the updated CFMM reserves are $(x',y').$ %We will refer to the amount of assets in the reserves as a CFMM's \textit{state}.
%The CFMM may charge an additional fee. 
The following characterization of trading functions is standard in the literature and is required to ensure that the CFMM works as intended.
\begin{assumption}
\label{ass:cfmm}
 CFMM trading functions are quasi-concave, continuous, and non-decreasing (in both coordinates) on $\mathbb{R}^2_+$.
\end{assumption}

This characterization implies that 
given any level curve of the trading function $f$,
the amount $x$ of asset $X$ in the reserves, uniquely determines the amount of $Y$, which we denote $\mathcal{Y}(x)$ 
%We denote it by $\mathcal{Y}(x)$ 
when the level curve of the function $f$ is clear from the context. 

Commonly studied CFMMs include the constant product $f(x,y)= xy$, used by, for example, the exchange Uniswap \cite{uniswapv2}, and $f(x,y) = 2- e^{-x} -  e^{-y}$ which implements the  Logarithmic Market Scoring Rule (LMSR) \cite{hanson2007logarithmic}.
%Below are some examples of commonly studied CFMMs.
%
\iffalse
\begin{example}[CFMM Trading Functions]
\label{ex:cfmms}
~
\begin{itemize}[leftmargin=10pt]
\item The exchange Uniswap \cite{uniswapv2}  uses $f(x,y)= xy$. % The spot price from $A$ to $B$ is $b/a$.
\item The Constant Sum CFMM uses $f(x,y) =  r x + y$ with $r >0.$
\item The Logarithmic Market Scoring Rule \cite{hanson2007logarithmic} corresponds to a trading function $f(x,y) = 2- e^{-x} -  e^{-y}$
	 \cite{univ3paradigm}.
\end{itemize}
\end{example}
\fi

CFMMs, in practice, charge a ``trading fee'' to all trades which is a percentage of the trade size. %Any fees that the CFMM charges may either be re-invested in its market-making reserves or may be taken by the liquidity provider. 
%CFMMs, in practice, charge a ``trading fee'' to all trades which is $c$ percentage of the trade size. %Any fees that the CFMM charges may either be re-invested in its market-making reserves or may be taken by the liquidity provider. 
For simplicity, we assume that any fee is not added to the CFMM reserves -- it is taken by the LP. 

An important quantity for the analysis of CFMMs is the price it provides for an infinitesimal trade, referred to as its spot price.

\begin{definition}[Spot Price]
\label{defn:spot}
%The {\it spot exchange rate} is the exchange rate for a trade of infinitesimal size. 
%
At state $(\hat x, \hat y),$ the spot price of a CFMM with trading function $f$ is
$\frac{\partial f}{\partial x}/\frac{\partial f}{\partial y}$ at $(\hat x, \hat y)$.

%The spot exchange rate may not be unique.  
When $f$ is not differentiable, the spot price is any subgradient of $f$.
%When $\hat x=0$, the spot price is $[\frac{\partial f}{\partial x}/\frac{\partial f}{\partial y}, \infty)$,
%and when $\hat y =0$, the spot price is $[0,\frac{\partial f}{\partial x}/\frac{\partial f}{\partial y}]$.
\end{definition}

\iffalse
\begin{definition}[Spot Price]
\label{defn:spot}
The {\it spot price} from asset $X$ to $Y$ for a CFMM with trading function $f$ at state $(\hat x,\hat y)$ is 
$\frac{\partial f}{\partial x}(\hat{x}, \hat y)/\frac{\partial f}{\partial y}(\hat{x}, \hat y)$.  
\end{definition}
\fi 
%When the spot price is set-valued, the CFMM, by construction, buys X at the lowest price in the set and sells X at the highest price. 
Assumption~\ref{ass:cfmm} implies the price offered to a trader is never better than the spot price. This has an important implication: when the price at the external market deviates, arbitragers align the CFMM's spot price with it. In this sense, a CFMM is a ``truthful'' mechanism -- that an arbitrager cannot do better by ``reporting'' a false price to the CFMM.
We make an additional assumption.

\begin{assumption} \label{ass:cfmm2}
    The CFMM never runs out of an asset.
\end{assumption}

\iffalse
\begin{observation} \label{obs:all-prices}
    By Assumptions~\ref{ass:cfmm} and~\ref{ass:cfmm2}, the CFMM ``makes'' a market in the entire price range $(0, \infty).$
\end{observation}
\fi

\subsection{Differential Privacy}
The traditional notion of DP is for a centralized model (``global'' DP), where a trusted data curator holds all users' accurate data and reveals aggregate statistics to queries with the constraint that it cannot be inverted to infer any specific user's data. We consider here a different notion of privacy, called personalized local differential privacy \cite{chen2016private}, wherein each user's data is privatized (usually by adding noise) \textit{before} being stored with the mechanism. The privacy guarantee is given by a parameter $\varepsilon$; higher $\varepsilon$ implies worse privacy protection. 

\iffalse
\begin{definition}[Local Differential Privacy \cite{kasiviswanathan2011can}] \label{def:ldp}
    A randomized algorithm $\mathcal{A}$ satisfies $\varepsilon-$LDP if for any pair of values $l,l' \in \tau,$ and for any measurable subset $O \subseteq Range(\mathcal{A}),$
    $$ Pr[\mathcal{A}(l) \in O] \leq \exp (\varepsilon) \cdot Pr[\mathcal{A}(l') \in O],$$
    where the probability space is over the randomness in $\mathcal{A}.$
\end{definition}
\fi

% \citet{chen2016private} proposed a generalization where each user can specify their privacy level $\varepsilon$ and masking interval $\tau.$

 \begin{definition}[Personalized Local Differential Privacy (PLDP)] \label{def:pldp}
  Given the personalized privacy specification $(\tau_i,\varepsilon_i)$ of user $i$,  a randomized algorithm $\mathcal{A}$ satisfies $(\tau_i,\varepsilon_i)-$PLDP for $i$ if for any pair of values $l,l' \in \tau_i,$ and for any measurable subset $O \subseteq Range(\mathcal{A}),$
    $$ Pr[\mathcal{A}(l) \in O] \leq \exp (\varepsilon_i) \cdot Pr[\mathcal{A}(l') \in O],$$
    where the probability space is over the randomness in $\mathcal{A}.$
\end{definition}
Observe that traditional local differential privacy (LDP) \cite{kasiviswanathan2011can}  is a special case of PLDP where all users have the same $\tau$ and $\epsilon.$

A larger masking interval $\tau$ implies stronger privacy, and within that interval, a smaller $\varepsilon$ corresponds to a better privacy guarantee. For example, $\varepsilon = 0$ corresponds to the case that all points in masking interval $\tau$ produce the same distribution of the algorithm's output. 

\section{Model}
We describe here the model of the adversary, our noisy CFMM mechanism, the external market price, and the associated arbitrage. 
%We first discuss the capabilities of the adversary.
\subsection{Adversary Attack Model} \label{sec:attack}
We adopt the following attack model, which was first given by \citet{angeris2021note}, and also used by \citet{chitra2022differential}.
\begin{enumerate}[leftmargin = 10pt]
\item Eavesdropper Eve knows the trading function of the CFMM.
%\geoff{and the function's value at the current reserves (the current level set)}.
%\mohak{They don't need to know that; point 4 helps with that.}
    \item Eve knows \textit{when} trader Tod makes a transaction with the CFMM.
    \item Eve can query the \textit{spot price} of CFMM.
    \item Eve can query whether a given non-zero trade is \textit{valid}.
    \item Eve can do (3) and (4) \textit{before and after} Tod's trade.
\end{enumerate}

\citet{angeris2021note} showed that this much information is sufficient for Eve to infer exactly the amount traded by Tod in a traditional CFMM.%, which is why we need a privacy protection mechanism for CFMMs.

\subsection{System Model}
We borrow here the model of some private smart contract systems
(such as Hawk \cite{kosba2016hawk}), which delegate private computation to a ``manager,'' that can be implemented via trusted execution environments (TEE) or multiparty computation (MPC) protocols, as in, e.g., \citet{banerjee2021zkhawk} and  \citet{baum2022eagle}.
We specifically require a way for the CFMM to maintain private state,
and to execute code on that private state.
This private information includes not only the CFMM reserves
but also a private account $\mathcal{H}$, used in the mechanism discussed below.  We also require a source of unpredictable (pseudo)randomness, such as the output of a VRF \cite{micali1999verifiable}.
%We assume that an external operator ensures that $\mathcal{H}$ never runs dry.

We also assume the existence of arbitrageurs,
who may trade on the CFMM and on external, highly liquid exchanges,
and that arbitrageurs are aware of the market price on external exchanges.

%
%However, we assume that the CFMM can maintain other private information not covered by this model, including ownership of additional assets in a private account.

\subsection{Privacy Protection Mechanism for CFMMs}
The noisy CFMM randomly distorts its state after every trade.

\underline{\textbf{Noisy CFMM Mechanism}}
\begin{enumerate}[leftmargin = 10pt]
\item The CFMM maintains a ``hidden'' private account $\mathcal{H}$ of reserves. This account is used to make ``noise trades''%\footnote{Not to be confused with the traditional finance notion of a noise trader who trades for intrinsic reasons and not in response to price movements.%; here we use this phrase to mean an actual noise trade executed by the CFMM with itself for privacy protection and adopt this nomenclature from \citet{frongillo2018bounded}.
%}''
as follows.
    \item At some point in time, the CFMM state is $(x, \mathcal{Y}(x)).$% has $x$ units of X and $\mathcal{Y}$ units of $Y$ in its reserves.
    \item Trader $i$ makes a trade selling $\Delta_i$ units of X to the CFMM and specifies their privacy requirements $(\tau_i, \varepsilon_i)$ such that $\Delta_i \in \tau_i.$ 
    \item Trader $i$ gets $\mathcal{Y}(x) - \mathcal{Y}(x + \Delta_i)$ units of $Y$ in exchange for $\Delta_i$ units of X, and pays a fee, which is the sum of a ``trading fee'' %\geoff{Trading fee comes out of nowhere} \mohak{added trading fee to preliminaries}
    and a  ``privacy fee'' $\gamma_i.$ The fee is not added to the CFMM state.
    \item The CFMM immediately makes a random trade of $\eta_i \sim \mathcal{D}_i$ units of $X$ with the hidden account $\mathcal{H}$.
    \item The new CFMM state is $x' = x + \Delta_i + \eta_i$, $y'=\mathcal{Y}(x')$ . This state can be queried by  the next trader (and Eve, via the attack of \S\ref{sec:attack}).
\end{enumerate}
We assume that an external operator supports the CFMM
to ensure that $\mathcal{H}$ never runs dry, but if
$\mathcal{H}$ cannot support the maximum possible noise for a trade request, the CFMM rejects the request.

%We further assume 
%that the hidden account $\mathcal{H}$ can be kept private via ZKPs and 
%that the CFMM operator ensures that $\mathcal{H}$ has enough funds to support the noise trades associated with reasonable sizes of next trade requests. When the reserves of $\mathcal{H}$ cannot support the maximum possible noise associated with a trade request, the CFMM automatically rejects the request\geoff{Someone might worry that this rejection leaks information, but not sure what else you can do.} Further implementation details are beyond the scope of this paper.\geoff{Maybe say future work? idk}

%The `trading fee' can be determined based on previous work \cite{evans2021optimal,angeris2021optimal,milionis2022automated}. 
%The noise trade distributions $\mathcal{D}_i$ and privacy fee $\gamma_i$ are this paper's main topics of study. 

We use noise trade distributions of bounded support -- we will give examples in \S~\ref{sec:dist}. 
We consider noise trade distributions which, for a given CFMM, depend only on the trade amount $\Delta_i$ and privacy specification $(\tau_i,\varepsilon_i).$ Importantly, it is independent of the CFMM state. This setting is standard in the DP literature.  
The privacy fee $\gamma_i$ depends on the CFMM state and the noise trade distribution $\mathcal{D}_i$. 

Since our mechanism must also supports non-private trades as traditional CFMMs do, the noise trade $\eta_i$ and the privacy fee $\gamma_i$ are zero when  $\varepsilon_i = \infty$ or $\tau_i = \{\Delta_i\}$ for user $i$.
\iffalse
\begin{observation}
   For user $i,$ if $\varepsilon_i = \infty$ or $\tau_i = \{\Delta_i\},$  the noise trade $\eta_i = 0$ and the privacy fee $\gamma_i = 0.$
\end{observation}
\fi
\subsection{Arbitrage and Additional Arbitrage}
In this paper, we study the case where an external market with infinite liquidity exists, and arbitrageurs can trade both with the CFMM and the external market. %However, not all traders may have this capability, for example, if  registered only on one exchange. 
We consider the external market price as the ``true'' price. %Arbitrageurs ensure that the spot price of the CFMM closely follows the true price. 
This setting is standard in DeFi literature on CFMMs, for example, in \citet{goyal2022finding, evans2021optimal, milionis2022automated}. 
%We believe that our analysis of arbitrage and fee is also important when the CFMM is the primary market. 
Our result also holds when arbitrageurs trade on private knowledge of future or true prices, which is important when a bigger external market does not exist. Whenever the CFMM spot price deviates from the true price, an arbitrageur makes a risk-less profit by aligning CFMM's spot price with the true price.
%Informally, our results only require that arbitrage is well-defined, which can also be true when traders have private knowledge of future asset valuations. However, formally extending our results to such a setting requires further investigation.

%Previous results have shown that CFMMs lose value due to arbitrage \cite{milionis2022automated}, and are compen

We now illustrate the \textit{additional} arbitrage problem with the noisy CFMM mechanism. 
Suppose for the moment that the privacy fee were zero and 
%Consider, for a moment, that the privacy fee is zero. 
that Tod has exclusive access to the CFMM. Tod can exploit the noisy CFMM mechanism in the following manner.
\subsubsection{Arbitrage on Noisy CFMM} \label{sec:arb}
\begin{enumerate} [leftmargin = 10pt]
    \item Tod observes the CFMM spot price $p$ and the true price $\hat p$.
    \item If $p\neq \hat p,$ they make a trade with the CFMM, such that the post-trade CFMM spot price $\tilde p$ satisfies $\tilde p = \hat p.$ They specify some non-trivial privacy requirements. Tod makes a risk-free profit by taking the reverse trade on the external market.
    \item The CFMM executes its noise trade and ends up with a spot price $p'$, which is not equal to $\hat{p}$ with nonzero probability.
    \item Steps 2 and 3 repeat until a zero noise trade is drawn.
    
\end{enumerate}

Observe that the arbitrage profit of Step 2 is not due to the 
privacy feature of the CFMM. In this paper, we do not account for this component of the CFMM's loss. %\geoff{Can we cite a couple examples of this being standard in the literature?}
%\mohak{I think this is borderline controversial and people who we cite may take offence. idk I may be overthinking. Anyway just adding it.}
In a traditional CFMM, arbitrage stops at Step 2. In a noisy CFMM, additional arbitrage profit is created in Step 3, which is the subject of study of this paper.

\begin{definition}[Additional Arbitrage]
    A CFMM $C$ and a noisy CFMM $\tilde C$ have identical states and trading functions. For any true price $\hat p$, the additional arbitrage is the difference of the maximum possible profit in expectation from trading with $\tilde C$ and $C$.
\end{definition}

\section{Truthfulness and Priceability of Noisy CFMM Mechanism}
The strategy above is not the only possible thing trader Tod might do. For example, they could conceivably leave the CFMM at a spot price $\tilde p$ different from the true price $\hat p$ and make a sequence of multiple trades with the noisy CFMM for a higher overall profit. This motivates the following property.

\begin{definition}[Truthfulness]\label{def:truthful}
    A noisy CFMM mechanism is truthful if a trader with no privacy requirements, strictly increasing utility in Y,  no utility for X,
    %\geoff{And no private information/utility for X, right?}
   % \mohak{we assume that the arbitrageur exactly knows the price of X}
     exclusive access to the CFMM, and access to the external market, always maximizes their utility by trading with the CFMM only once such that the post-trade spot price is equal to the true price of X, and with privacy level $\varepsilon = \infty$.
\end{definition}

This definition does not apply to traders who trade for intrinsic reasons without regard to price movements (so-called ``uninformed traders''). It also does not apply to traders with privacy requirements since we do not quantify the value that traders attach to privacy protection. It applies to strategic and informed traders who perform arbitrage without privacy requirements. Although defined  narrowly, our notion of truthfulness is important for two reasons: (1) it explains if the noisy CFMM will closely follow the true price and therefore be attractive to uninformed traders, and (2) it paves the way for us to compute correct privacy fees and give a minimal characterization of acceptable noise trade distributions.

We define a property to study the economic feasibility of the privacy feature on CFMMs.

\begin{definition}[Priceability] \label{def:priceability}
    A noisy CFMM mechanism is priceable if there always exists a privacy fee which depends only on the CFMM state, its trading function, and the noise trade distribution such that the mechanism has the following property.  
    
    For any true price and CFMM state, the additional arbitrage is zero.
\end{definition}

The crucial feature of Definition ~\ref{def:priceability}
is that the privacy fee cannot depend on the true price $\hat p.$ Observe that for a privacy fee of zero, the noisy CFMM creates strictly greater arbitrage profit than a traditional CFMM  when the CFMM has non-zero curvature (\S\ref{sec:arb}). %Therefore, the correct privacy fee is strictly positive.

Truthfulness and priceability have different motivations. 
Truthfulness ensures that the CFMM spot price follows the true price and is attractive to uninformed traders.  
Priceability ensures that the CFMM does not lose money to additional arbitrage created by the privacy feature. 
However, these two properties are closely related technically. The following observations follow from the definitions of these properties.

\begin{observation} \label{obs:truthful-implies-priceable}
    Truthfulness implies priceability.
\end{observation}
\begin{observation}
  There exists a privacy fee function under which a priceable noisy CFMM mechanism is truthful.
\end{observation}
\iffalse
\begin{proof}
    Follows from Definitions~\ref{def:truthful} and~\ref{def:priceability}.
\end{proof}
\fi
%Recall that the noise trade distribution is a function of the trade amount $\Delta,$ privacy level $\varepsilon,$ and masking interval $\tau.$ 

Before discussing our results characterizing the noise trade distribution and privacy fee, we first give an alternate formulation of a CFMM.
A level curve of a trading function can be seen as a map from the amount of X in the reserves to the CFMM spot price. 
We denote this map by $P(x)$, where the trading function and the level curve are clear from the context. When the trading function is not differentiable,
$P(x)$ is the largest subgradient at that point. 
Assumptions ~\ref{ass:cfmm} and ~\ref{ass:cfmm2} imply that $P(x)$ has the following properties.
%has the following properties implied by Assumptions~\ref{ass:cfmm} and~\ref{ass:cfmm2}.

\begin{observation}
   Spot price $P(x)$ is monotonically decreasing, $P(x) \rightarrow \infty$ as $x \rightarrow 0$ and $P(x) \rightarrow 0$ as $x \rightarrow \infty.$
\end{observation}

We similarly define $P^{-1} (p)$ as the largest $x$ for which the spot price is $p$ when the trading function and its level curve are clear from the context (since we do not re-invest fee into the reserves, our CFMM is path independent and stays at the initial level curve).
We make the following assumption for technical ease.
\begin{assumption}
    The trading fee is zero.
\end{assumption}

This assumption is for expository clarity, and is not \textit{required}.
%This assumption is not \textit{required}, but helps to convey them clearly. 
When a $c$ percent trading fee is charged, the truthfulness property becomes \textit{approximate} such that an arbitrageur corrects the CFMM spot price only if it is more than $c$ percent away from the true price. In practice, $c$ is generally below $1\%.$
Coming to our results, we first describe what does not work for truthfulness.

\begin{theorem} \label{thm:non-zero-not-truthful}
    If the noise trade distribution $\mathcal{D} (\Delta, \varepsilon, \tau)$ has non-zero-mean for some $(\Delta, \varepsilon, \tau)$, then the noisy CFMM mechanism is not truthful for any finite privacy fee.
\end{theorem}

\begin{proof}
Denote the trade size, privacy level, and masking interval for which $\mathcal{D}$ has non-zero mean by $(\tilde \Delta, \tilde \varepsilon, \tilde \tau).$ Denote the mean by $\mu.$
We consider two cases where $\mu$ is either positive or negative.

\textbf{Case 1:} $\mu >0.$
Here, the situation where truthfulness is violated has the true price $\hat p$ exceeds the initial spot price $p.$

%Denote the percentage trading fee by $c.$ 
The arbitrageur's profit under the truthful strategy is:
\begin{equation} \label{eq:truthful-profit}
     \int_{P^{-1}(p)}^{P^{-1}(\hat p)} (P(a) - \hat p) ~da.
\end{equation}

Consider the following strategy for the arbitrageur:
\begin{enumerate}[leftmargin = 15pt]
    \item Make a trade of $(\tilde \Delta, \tilde \varepsilon, \tilde \tau),$ and pay the privacy fee $\gamma.$
    \item Make a non-private trade with post-trade CFMM spot price $\hat p.$
\end{enumerate}

When the noise trade drawn is $\eta,$ the profit with this strategy is:
$$ - \gamma + \int_{P^{-1}(p)}^{P^{-1}(p) + \tilde \Delta} (P(a) - \hat p) ~da +  \int_{P^{-1}(p) + \tilde \Delta + \eta }^{P^{-1}(\hat p)} ( P(a) - \hat p) ~da  $$

The excess gain from deviating from the truthful strategy is 
\begin{align*}
    &- \gamma + \int_{P^{-1}(p) + \tilde \Delta + \eta}^{P^{-1}(p) + \tilde \Delta } (P(a) - \hat p) ~da \\
   % = &- \gamma 
   % + \int_{P^{-1}(p) + \tilde \Delta+ \eta}^{P^{-1}(p) - \tilde \Delta}  P(a) ~da 
   % + \int_{P^{-1}(p) + \tilde \Delta +\eta}^{P^{-1}(p) + \tilde \Delta }  \hat p ~da \\
     = &- \gamma 
    + \int_{P^{-1}(p) + \tilde \Delta + \eta}^{P^{-1}(p) + \tilde \Delta}  P(a) ~da 
    + \eta  \hat p
\end{align*}
Recall that $\mathbb{E}(\eta) = \mu > 0$ in this case.
The first two terms are bounded and independent of $\hat p$. For large enough $\hat p$, the positive third term dominates, and the result follows for this case.
%We now move on to the case where $\mu <0.$

\textbf{Case 2:} $\mu <0.$
Here we construct a scenario where the true price $\hat p$ is less than 
%\geoff{is subceed a word?} 
the initial spot price $p.$
The arbitrageur's profit under the truthful strategy is same as \eqref{eq:truthful-profit}.
%$$  \int_{P^{-1}(p)}^{P^{-1}(\hat p)} ( P(a) - \hat p) ~da.$$

\iffalse
\geoff{Shouldn't this profit be max(expression, 0)?  
The arbitrageur might reasonably choose to deviate
by not trading if fees are too high.

Or rather, this expression is the truthful strategy,
but I think we should emphasize that the nontruthfulness
of the strategies you construct are coming from extra arb opportunities, not from charging high trading fees.}

\geoff{
Why are the expressions for truthful profit different in the two cases?  Something doesn't quite add up to me.
}
\fi

Consider the following three-step strategy for the arbitrageur:
\begin{enumerate} [leftmargin=15pt]
\item Make a non-private trade with the CFMM to buy X such that its spot price becomes $p'$ (where $p'$ is a large value and will be set precisely later).
    \item Make a trade of $(\tilde \Delta, \tilde \varepsilon, \tilde \tau),$ and pay the privacy fee $\gamma.$
    \item Make a non-private trade with post-trade CFMM spot price  $\hat p.$
\end{enumerate}

When the noise trade drawn is $\eta,$ the profit with this strategy is:
\begin{align*}
     &- \gamma + \int_{P^{-1}(p)}^{P^{-1}(p')} ( P(a) - \hat p) ~da +  \int_{P^{-1}(p') }^{P^{-1}(p') + \tilde \Delta} ( P(a) - \hat p) ~da \\
     &+   \int_{P^{-1}(p') + \tilde \Delta +\eta }^{P^{-1}(\hat p)} ( P(a) - \hat p) ~da 
\end{align*}

The excess gain from deviating from the truthful strategy is: 
\begin{align*}
    &- \gamma + \int_{P^{-1}(p)}^{P^{-1}(p')} ( P(a) - \hat p) ~da +  \int_{P^{-1}(p') }^{P^{-1}(p') + \tilde \Delta} ( P(a) - \hat p) ~da \\
     &+   \int_{P^{-1}(p') + \tilde \Delta +\eta }^{P^{-1}(\hat p)} ( P(a) - \hat p) ~da - \int_{P^{-1}(p)}^{P^{-1}(\hat p)} ( P(a) - \hat p) ~da\\
     = &- \gamma + \int_{P^{-1}(p') + \tilde \Delta +\eta}^{P^{-1}(p') + \tilde \Delta} ( P(a) - \hat p) ~da\\
= &- \gamma + \eta \hat p +  \int_{P^{-1}(p') + \tilde \Delta +\eta}^{P^{-1}(p') + \tilde \Delta} P(a)  ~da
\intertext{Since $P(\cdot)$ is monotonically decreasing and positive,}
\geq &- \gamma + \eta \hat p -   \eta P(P^{-1}(p') + \tilde \Delta)
\end{align*}

Recall that $\mathbb{E}(\eta) = \mu <0.$
We construct the case where the true price $\hat p$ is smaller than $\frac{1}{|\mu|}$.  We set $p'$ in the arbitrageur's strategy such that $-\mu   P(P^{-1}(p') + \tilde \Delta)$ is larger than $- \gamma + \mu \hat p.$ Such a $p'$ always exists under Assumptions~\ref{ass:cfmm} and~\ref{ass:cfmm2}. Since our noise trade distributions have bounded support, a large enough level curve of the CFMM exists where the noise trade is supported at spot price $p'.$  
This implies that, in expectation, the arbitrageur can obtain a larger profit with our 3-step strategy than under the truthful strategy. This completes the proof.

Discussion: We need two different arbitrage strategies for $\mu >0$ and $\mu < 0$ cases due to an asymmetry between X and Y. The noise trades are in units of $X$, and when the true price of $X$ is small, the loss of these noise trades will be small when made in a state close to the true price. To show that the arbitrageur can nonetheless make a large profit, we need an extra step in their strategy.
%This completes the proof. \mohak{explain why we need different cases succinctly.}
\end{proof}

This result shows that the noise trade distribution being zero-mean is necessary for truthfulness. For non-zero-mean noise trade distributions, the proof shows that there exists a true price $\hat p$ under which the arbitrageur is better off paying the privacy fee in exchange for the additional arbitrage that the noise trade creates. %In this situation, we say that the mechanism is not ``priceable.'' Formally: %This means that the mechanism designer 
Also note the following corollary.% of Theorem~\ref{thm:non-zero-not-truthful}.
\begin{corollary}
  If the noise trade distribution $\mathcal{D} (\Delta, \varepsilon, \tau)$ is not zero-mean for some $(\Delta, \varepsilon, \tau)$, the noisy CFMM is not priceable.
  \end{corollary}
  \iffalse
\begin{proof}
Observe from the proof of Theorem~\ref{thm:non-zero-not-truthful} that the additional payoff of an arbitrageur in a noisy CFMM above that in a simple CFMM can be unbounded. %Therefore no finite privacy fee can account for it.
%    Since the mechanism is not truthful for any finite privacy fee, an arbitrageur has a strategy which pays off an unbounded amount of Y in expectation for some true price of X. Since the payoff of the truthful reporting strategy is finite, the mechanism cannot be priceable.
\end{proof}
\fi
This result establishes that the noise trade distribution being zero-mean is necessary for priceability too. We later show that it is also sufficient (Corollary~\ref{cor:priceable}).
Before discussing the result, we define a quantity termed `noise fee' as a function of the CFMM state, trading function, and noise trade distribution. It captures the expected arbitrage gain from making the trade that ``reverses the noise'' when the CFMM spot price is the true price.

\begin{definition}[Noise fee]\label{def:noise-fee}
    For CFMM state $(\tilde x,\mathcal{Y}(\tilde x)),$ trade $\tilde \Delta$, privacy specification $(\tilde \tau, \tilde \varepsilon)$ and noise trade distribution $\mathcal{D},$ the noise fee $\Gamma$ is:
    $\Gamma = \int_{-\infty}^{\infty} \mathcal{D}(\eta) \int_{\tilde x + \tilde \Delta + \eta}^{\tilde x + \tilde \Delta} (P(a) - P(\tilde x + \tilde \Delta)) ~da ~d\eta.$

    For zero-mean noise trade distributions, the noise fee simplifies to $\Gamma = \int_{-\infty}^{\infty} \mathcal{D}(\eta) \int_{\tilde x + \tilde \Delta + \eta}^{\tilde x + \tilde \Delta}  P(a) ~da ~d\eta$.    
\end{definition}

\iffalse
\begin{observation} \label{obs:fee-zero-mean}
For zero-mean noise trade distributions, the noise fee simplifies to $\Gamma = \int_{-\infty}^{\infty} \mathcal{D}(\eta) \int_{\tilde x + \tilde \Delta + \eta}^{\tilde x + \tilde \Delta}  P(a) ~da ~d\eta$.    
\end{observation}
\fi

See that the noise fee is defined with the idea that it must account for the additional arbitrage \textit{when} the post-trade spot price is equal to the true price. Importantly, it is oblivious to the actual true price of X. However, as we show in the next result (Theorem~\ref{thm:truthful}), for zero-mean $\mathcal{D},$ the noise fee is the ``right'' choice for the privacy fee and it makes the noisy CFMM mechanism truthful.

\begin{theorem}\label{thm:truthful}
If the noise trade distribution is zero-mean, a privacy fee exists under which the noisy CFMM mechanism is truthful.

The minimum such privacy fee is the noise fee $\Gamma$ (Definition~\ref{def:noise-fee}).
\end{theorem}

\iffalse
\geoff{I don't understand this proof at all.

Why are there $n$ trades? I think the intent is to say
something like "for any strategy/sequence of n trades"
but an arbitrageur could conceivably adapt their trades
in response to observed noise.

Something around the framing needs to be fixed, I think.
But also the integrals suddenly change indexing,
and it's not clear what's going on.}
\fi

Appendix~\ref{app:proof-truthful} gives a proof, which shows that for any finitely-bounded sequence of trades made by an arbitrageur, even adaptive to the noise, the expected gain in excess of that of the truthful strategy is equal to the noise fee. The proof invokes the Doob's optional stopping theorem on martingales \cite[Chapter 12.5]{grimmett2020probability}.
Thus, risk-neutral arbitrageurs should treat a noisy CFMM the same as a traditional CFMM. Privacy therefore does not hamper the basic functionality of a CFMM of offering close to true prices in the presence of arbitrageurs.

The following corollary follows from Observation~\ref{obs:truthful-implies-priceable}.
\begin{corollary} \label{cor:priceable}
       A noisy CFMM mechanism is priceable if the noise trade distribution $\mathcal{D}$ is always zero-mean.
\end{corollary} 

Therefore, for zero-mean noise trade distributions, the mechanism can protect the CFMM from additional arbitrage \textit{without observing} the true price. The privacy feature does not harm the returns of risk-neutral LPs when the privacy fee is set correctly. %Priceability gives that such a privacy fee exists, and by Theorem~\ref{thm:truthful}, the noise fee (Defnition~\ref{def:noise-fee}) is the minimum fee that achieves this property.
\subsection{Choice of Noise Trade Distributions}\label{sec:dist}
We first discuss a noise distribution from the LDP literature, which is well-suited for our application. %We first give a distribution with support on only two points.
%\mohak{Need clarity in this section; everything about optimality is hand wavy. Give an example, and also mention clearly what we mean by divergence and optimal.}
\begin{definition}[Binary Mechanism of \citet{duchi2018minimax}]
    For privacy specification $(\tau = [l,u],\varepsilon)$ and data (trade) $\Delta,$ add noise
    \begin{enumerate}
        \item $(\frac{u+l}{2} - \Delta - \frac{u-l}{2} \frac{e^{\varepsilon} +1}{e^{\varepsilon} -1} )$ with probability $\frac{1}{2}[1-\Delta' \frac{e^{\varepsilon} -1}{e^{\varepsilon} +1}].$
        \item $(\frac{u+l}{2} - \Delta + \frac{u-l}{2} \frac{e^{\varepsilon} +1}{e^{\varepsilon} -1} )$ with probability $\frac{1}{2}[1+\Delta' \frac{e^{\varepsilon} -1}{e^{\varepsilon} +1}].$
    \end{enumerate}
    Here $\Delta' = \frac{2\Delta - (u+l)}{u-l}.$
\end{definition}

Importantly, the noise is zero-mean and of bounded magnitude. %unlike the well-studied Laplace mechanism \cite{dwork2014algorithmic}.
%\citet{duchi2018minimax} showed that this noise is zero-mean. 
\citet{kairouz2014extremal} showed that in the high privacy regime, i.e., for all $\varepsilon$ less than some $\varepsilon^*,$ the binary mechanism is optimal for minimizing the `information loss' due to privacy.

%They also generalized their result to maximizing all divergences. In our context, their result implies that the binary mechanism minimizes the privacy fee $\Gamma$ in the high privacy regime.

%Beyond the high privacy regime, \citet{wang2019collecting} gave a `piecewise mechanism' (PM) with a noise distribution with three flat regions over a bounded support. %They showed that PM outperforms the binary mechanism for large $\varepsilon$ for minimizing noise variance. 

We are interested in designing noise distributions that minimize the privacy fee. Observe that the privacy fee is a convex (linear) function of $\mathcal{D}(\eta)$ for all $\eta$ for all state $\tilde x$ and trade $\tilde \Delta$. %This follows from the fact that the derivative is $-P(\tilde x + \tilde \Delta + \eta)$, and the second derivative is positive since $P(\cdot)$ is decreasing.
The privacy requirements can be expressed as convex constraints in $\mathcal{D}(\eta)$, as in \citet{kairouz2014extremal}. The mean being zero is also a linear constraint. Given a state and trading function, we can give a convex program to find the ``cheapest'' noise trade distribution that preserves $(\tau, \varepsilon)$-PLDP. However, since noise distributions have to be independent of the CFMM state, finding a distribution that works reasonably well for all states is not obvious and is the subject of future work.

%The noise variance closely approximates the privacy fee for the constant product CFMM.
%\mohak{either explain it or omit it.}

\subsection{Relation Between Privacy Fee and Liquidity}
%\mohak{Needs a lot more high level picture}
Notice from Definition~\ref{def:noise-fee} that the privacy fee is zero in a region where the spot price is constant. This matches the intuition that the constant sum CFMM provides privacy for free due to its lack of curvature -- it has ``infinite'' liquidity at a price point. We define liquidity at a price as follows.

\begin{definition}[Liquidity] \label{def:liquidity}
When spot price $P(\cdot)$ is differentiable and $\frac{dP(x)}{dx} \vert_{x=P^{-1}(p)} \neq 0,$ the liquidity
    $L(p)$ is $\left(1/\frac{dP(x)}{dx}\right) \Big\vert_{x= P^{-1}(p)}$.
    
     $L(p) = 0$ when $P(\cdot)$ is not differentiable, and
     
      $L(p)$ is undefined when $\frac{dP(x)}{dx}\Big\vert_{x= P^{-1}(p)}=0.$ 
\end{definition}

For a given trade, the privacy fee captures the additional arbitrage profit that can be captured by reversing the associated noise trade. Intuitively, this quantity is larger if the noise trade creates a larger ``impact'' on the CFMM spot price. This impact is larger when the liquidity is smaller. We illustrate this intuition with an example of the widely adopted constant product ``Uniswap'' CFMM and the binary mechanism in Appendix~\ref{sec:fee-uniswap}. This intuition holds more generally and formally, as in the following result.

%For $f(x,y) = xy$:  the spot price $P(x)$ is $ \frac{y}{x} = \frac{K}{x^2},$ where $K$ specifies the level curve of $f$. For initial state $\tilde x$ and trade $\tilde \Delta,$
\iffalse
\begin{align*}
\Gamma &=  \int_{-\infty}^{\infty} \mathcal{D}(\eta) \int_{\tilde x+ \tilde \Delta +\eta}^{\tilde x + \tilde \Delta}  \frac{K}{a^2} ~da ~d\eta = -\int_{-\infty}^{\infty}   \frac{\mathcal{D}(\eta) \eta K}{(\tilde x + \tilde \Delta)(\tilde x + \tilde \Delta+\eta)} ~d\eta.
\end{align*}
\fi
%Observe that doubling the liquidity at all prices will double $K$ and $\tilde x$ where $\tilde x$ is $P^{-1}(p)$ for spot price $p$. For trade $\tilde \Delta$ and distribution $\mathcal{D}$, the fee $\Gamma$ will approximately \textit{halve}, the approximation being closer for larger $\tilde x$. The result holds more generally.

\begin{proposition}\label{thm:liq-fee}
 When the noise trade distribution is zero-mean, the privacy fee for a trade $\tilde \Delta$ at spot price $p$ is approximately inversely proportional to the CFMM liquidity at $p$ when $\tilde \Delta$ and the maximum possible size of the noise trade are $o(P^{-1} (p)).$
\end{proposition}
\begin{proof}
The privacy fee equals the noise fee by Theorem~\ref{thm:truthful}. We take a first-order approximation of the inverse of the liquidity.
In Definition~\ref{def:noise-fee} of the noise fee,
%is $\int_{-\infty}^{\infty} \mathcal{D}(\eta) \int_{P^{-1} (p) + \tilde \Delta + \eta}^{P^{-1} (p) + \tilde \Delta}  (P(a) - P(P^{-1} (p) + \tilde \Delta)) ~da ~d\eta.$  
the price difference $(P(a) - P(P^{-1} (p) + \tilde \Delta))$ in the interval $a \in [ P(P^{-1} (p)) + \tilde \Delta +\eta,  P(P^{-1} (p)) + \tilde \Delta] $ is approximately inversely proportional to the liquidity at price $p$ when the trade $\tilde \Delta$ and noise trade $\eta$ are $o((P^{-1} (p)))$.
\end{proof}

%This result holds more generally. 

%For simplicity, we discuss it for binary mechanisms. The noise values are $\eta_1<0$ and $\eta_2>0.$ The privacy fee per Definition~\ref{def:noise-fee} is:
%$ Pr[\eta=\eta_1] \int_{\tilde x + \tilde \Delta + \eta_1}^{\tilde x + \tilde \Delta}  P(a) ~da - Pr[\eta=\eta_2] \int_{\tilde x + \tilde \Delta}^{\tilde x + \tilde \Delta + \eta_2}  P(a) ~da   $. 

%Informally, the privacy fee increases with the ``price difference'' between the region $[\tilde x + \tilde \Delta + \eta_1, \tilde x + \tilde \Delta]$ and $[\tilde x + \tilde \Delta, \tilde x + \tilde \Delta + \eta_2].$ Since $\eta$ is bounded, for large $\tilde x,$ this ``price difference'' is closely approximated by the inverse of the liquidity.

This result shows that more liquid CFMMs are better suited for the needs of privacy-seeking traders. However, these CFMMs are prone to larger divergence loss, and also require larger capital investment from LPs. A holistic analysis of the design trade-offs of a noisy CFMM mechanism, possibly with custom-made trading functions, is the subject of future work.

\iffalse
\mohak{

Let noise added be $n$. The post-trade state is $S = (X_t, Y_t)$. The post-privacy state is then $S' = ( X_t + n, Y_t + n').$ Since $n'$ is a function of $n$, we describe all results in terms of $n$. Spot price at $S$ is $p.$ Say $p$ is also the `true price'. Define the price as a function of $X$ as $P(X).$ Now, the money that can be made by an arbitrageur for a realized value of $n$ is $$ Arb(n) = \int_{x=X_t}^{X_t + n} (p -  P(x)) ~dx.$$

Taking expectation over noise $n$:

$$  \mathbb{E}[Arb] = \int_{-\infty}^{\infty} Q(n)  \int_{x=X_t}^{X_t + n} (p -  P(x)) ~dx~dn $$

This is precisely the amount that must be charged to the trader.

Working out examples:

For Uniswap CFMM: $p = \frac{Y}{X} = \frac{K}{X^2}.$

\begin{align*}
Arb(n) &= \int_{x=X_t}^{X_t + n} (\frac{K}{X_t^2} -  \frac{K}{x^2}) ~dx \\
\intertext{Change of Variable, $y = x - X_t$}
&= \int_{y=0}^{n} (\frac{K}{X_t^2} -  \frac{K}{(y+X_t)^2}) ~dy \\
&= p \frac{n^2}{n+x}.
%&= \int_{y=0}^{n} \frac{K((y+X_t)^2 - X_t^2)}{X_t^2 (y+X_t)^2 } ~dy \\
%&=   p \int_{y=0}^{n} \frac{(y^2+2y X_t)}{(y+X_t)^2 } ~dy \\
%&= p \int_{y=0}^{n} \frac{(\frac{y}{X_t})^2 + 2\frac{y}{X_t}}{(\frac{y}{X_t}+1)^2 } ~dy\\
%\intertext{Change of Variable, $z = \frac{y}{X_t}$}
%&= p X_t \int_{z=0}^{\frac{n}{X_t}} \frac{z^2 + 2z}{(z+1)^2 } ~dz
\end{align*}
}
\fi

 %\subsection{Discussion of Price Volatility}
\section{Conclusions} 
In this work, we develop a noisy CFMM mechanism for privacy protection.  %which randomly distorts its state after each trade. 
Users can specify their desired privacy level and a masking interval for their trade and pay a personalized privacy fee. We find the minimum fee required to ensure that arbitrageurs cannot exploit the CFMM privacy feature. We also show that the noise being zero-mean in trade size is a necessary and sufficient condition for such a minimal fee to exist (we call this condition the priceability of the mechanism). We also show that our noisy CFMM is truthful under this fee, i.e., arbitrageurs are incentivized to align the CFMM's spot price with the external market price. 
For future work, it would be useful to design the noise trade distributions which can reduce the cost implied by the privacy fee. Efficient instantiations of our model, particularly the requirement of a private, secure account for conducting the noise trades are also avenues for future research.
%For future work, it would be useful to generalize our results to a case where the CFMM is the only (or primary) market in the asset it trades, and the price dynamics are endogenous.   %We also studied the price volatility via numerical simulations and found that the privacy-preservation feature does not cause much additional price volatility.

%\textbf{Future Work}
%It would be useful to generalize our results to a case where the CFMM is the only (or primary) market in the asset it trades. In this case, the price distortion on the CFMM due to the privacy feature may have important downsides. %Tangentially related, batch exchanges provide a useful framework for fair pricing \cite{budish2015high, canidio2023arbitrageurs} and are also capable of providing privacy to traders \cite{zswap}. A theoretical analysis of the privacy properties of batch exchanges is a crucial direction for future work in DeFi privacy.

\bibliographystyle{ACM-Reference-Format}
\balance
\bibliography{ref}

\appendix
\section{Proof of Theorem~\ref{thm:truthful}} \label{app:proof-truthful}
\begin{proof}
%\mohak{Write more details to show that the strategy can be adaptive, $n$ may as well be unbounded.}

The arbitrageur may deploy a strategy of making a sequence of trades with the CFMM, and their strategy may be adaptive to the noise drawings on previous trades.

Say the arbitrageur makes a bounded sequence $S$ of trades with the noisy CFMM. Denote the post-trade amounts of X in the CFMM reserves by $s_i$ for trade $i \in S.$ $\eta_i$ denotes the noise trade after trade $i$. $s_i$ may depend on $\eta_j$ and $s_j$ for all trades $j$ done before trade $i$.  

Let the true price be $\hat p$, and the initial spot price be $p$. %In the optimal arbitrage strategy, $s_ = P^{-1}(\hat p),$ since otherwise, the arbitrageur could increase their profit by moving the CFMM to spot price $\hat p.$ Also, the final trade is made without privacy-requirement, since otherwise,
   The profit of the truthful strategy is:
$$ \int_{P^{-1}(p)}^{P^{-1}(\hat p)} (P(a) -  \hat p) ~da.$$

Denote $P^{-1}(p)$ by $s_0$ by and $\eta_0 = 0$ for notational convenience.
The profit of the alternate strategy is:
$$ -\sum_{i \in S} \gamma_i 
%+  \int_{P^{-1}(p)}^{s_1} (P(a) -  \hat p) ~da
+ \sum_{i \in S}~\int_{s_{i-1} + \eta_{i-1}}^{s_{i}} (P(a) -  \hat p) ~da.$$
%Here $\eta_i$ is the noise added by the CFMM after the $i$-th trade.
$\gamma_i$ is the privacy fee corresponding to trade $i$.

The excess profit over the truthful strategy is:
\begin{align*}
& -\sum_{i \in S} \gamma_i 
%+  \int_{P^{-1}(p)}^{s_1} (P(a) -  \hat p) ~da
+ \sum_{i \in S}~\int_{s_{i-1} + \eta_{i-1}}^{s_{i}} (P(a) -  \hat p) ~da - \int_{P^{-1}(p)}^{P^{-1}(\hat p)} (P(a) -  \hat p) ~da.
\intertext{This profit is less than the case the arbitrageur leaves the CFMM at spot price $\hat p,$ since if not, then another trade which moves the spot price to $\hat p$ makes non-zero profit. Further, since the integrands are all the same, we consider the remaining terms only, }
    \leq &-\sum_{i \in S} \gamma_i + \sum_{i \in S} \int_{s_{i} + \eta_{i}}^{s_{i}} (P(a) -  \hat p) ~da.\\
    =&-\sum_{i \in S} \gamma_i + \sum_{i \in S} \int_{s_{i} + \eta_{i}}^{s_{i}} P(a) - \sum_{i \in S} \int_{s_{i} + \eta_{i}}^{s_{i}} \hat p ~da.\\
   %\intertext{Since the noise trades are zero-mean, the $-\hat p$ terms are }
   % =&-\sum_{i \in S} \gamma_i + \sum_{i \in S} \int_{s_i + \eta_i}^{s_{i}} P(a) ~da.
    %\intertext{Taking expectation over the noise. The last term is  zero in expectation for zero-mean noise. Since all noise trades are independent,}
    %=&-\sum_{i \in S} \gamma_i + \sum_{i \in S} \int_{-\infty}^{\infty} \mathcal{D}_i(\eta_i) \int_{s_i + \eta_i}^{s_{i}} P(a) ~da.
\end{align*}
%\geoff{All the noise trades being independent is NOT TRUE if strategy is adaptive}
%The last step follows since the noise trades are zero-mean.
%\geoff{Is part of the expression here missing?  There at least needs to be an expectation somewhere.}
For each trade $i,$ $\Gamma_i = \mathbb{E}(\int_{s_{i} + \eta_{i}}^{s_{i}} P(a) -  \int_{s_{i} + \eta_{i}}^{s_{i}} \hat p ~da).$ This follows by definition of $\Gamma_i$ and the fact that $\mathbb{E}(\int_{s_{i} + \eta_{i}}^{s_{i}} \hat p ~da) = \hat p \mathbb{E}(\eta_i) =  0.$
Therefore the process of the arbitrageur's profit forms a martingle with expected values of the increments equal to $0$. For any bounded trade sequence, the Doob's optional stopping theorem \cite[Chapter 12.5]{grimmett2020probability} states that the expected value of the martingale is equal to its initial expected value.

Therefore setting $\gamma_i = \Gamma_i$ ensures that the truthfulness of the mechanism. This is also minimal, since if $\gamma_i < \Gamma_i,$ the arbitrageur can do better than the truthful strategy in only one step in expectation.
\end{proof}

\section{Privacy Fee on the constant product CFMM} \label{sec:fee-uniswap}
For $f(x,y) = xy$:  the spot price $P(x)$ is $ \frac{y}{x} = \frac{K}{x^2},$ where $K$ specifies the level curve of $f$. For initial state $\tilde x$ and trade $\tilde \Delta,$ the privacy fee is given by:
\begin{align*}
\Gamma &=  \int_{-\infty}^{\infty} \mathcal{D}(\eta) \int_{\tilde x+ \tilde \Delta +\eta}^{\tilde x + \tilde \Delta}  \frac{K}{a^2} ~da ~d\eta = -\int_{-\infty}^{\infty}   \frac{\mathcal{D}(\eta) \eta K}{(\tilde x + \tilde \Delta)(\tilde x + \tilde \Delta+\eta)} ~d\eta.
\end{align*}
%Observe that doubling the liquidity at all prices will double $K$ and $\tilde x$ where $\tilde x$ is $P^{-1}(p)$ for spot price $p$. For trade $\tilde \Delta$ and distribution $\mathcal{D}$, the fee $\Gamma$ will approximately \textit{halve}, the approximation being closer for larger $\tilde x$. The result holds more generally.

%Consider that the noise trade distribution is per the binary mechanism of Definition~\ref{def:binary}. 
Say, for the given privacy specification, the noise $\eta$ takes value $\eta_1$ with probability $p_1$ and $\eta_2$ with probability $p_2$. By the zero mean condition, we have $p_1\eta_1 + p_2 \eta_2 = 0.$ We have the privacy fee:
\begin{align*}
\Gamma &= \frac{-K}{(\tilde x + \tilde \Delta)}\left(\frac{p_1 \eta_1}{\tilde x + \tilde \Delta+\eta_1} + \frac{p_2 \eta_2}{\tilde x + \tilde \Delta+\eta_2} \right)\\
&=  \frac{-K}{(\tilde x + \tilde \Delta)}\left(\frac{p_1 \eta_1 (\tilde x + \tilde \Delta+\eta_2) + p_2 \eta_2(\tilde x + \tilde \Delta+\eta_1)}{(\tilde x + \tilde \Delta+\eta_1)(\tilde x + \tilde \Delta+\eta_2)}  \right)
\intertext{By the zero mean condition and since $p_1 + p_2 =1$ we have,}
\Gamma &=  \frac{-K\eta_1 \eta_2}{(\tilde x + \tilde \Delta)(\tilde x + \tilde \Delta+\eta_1)(\tilde x + \tilde \Delta+\eta_2)}. 
\end{align*}

Observe that doubling the liquidity at all prices will quadruple $K$ and double $\tilde x$ where $\tilde x$ is $P^{-1}(p)$ for spot price $p$. For trade $\tilde \Delta$ much smaller than $\tilde x$, that is, of order $o(\tilde x)$  and maximum noise magnitude $\max(|\eta_1|, |\eta_2|) $ of order $o(\tilde x)$, the fee $\Gamma$ will approximately \textit{halve} on doubling the liquidity. The approximation is closer for larger $\tilde x$. %A numerical example is in the Appendix.

Consider the following numerical example of the privacy fee for a typical trade size. Let the trade be $\tilde \Delta = 1.$ The privacy requirements are $\tau = [0,2]$ and $\varepsilon = 2.$ Let the trade be made when the CFMM spot price is $1$. Let the pool have $100$ units of X in this state. The privacy fee for this trade of 1 unit of X is $1.67 \times 10^{-2}$ units of Y. If the pool liquidity were twice as much, that is, if it had $200$ units of X at spot price $1$, then the privacy fee for this trade of 1 unit of $X$ would be $0.849\times 10^{-2}$ units of Y. Of course, the binary mechanism is not optimized for minimizing the privacy fee, and we can likely obtain much smaller fee for different noise distributions.
\end{document}